\documentclass[journal,twoside]{IEEEtran}
\IEEEoverridecommandlockouts
\usepackage{cite}
\usepackage{amsmath,amssymb,amsfonts}
\usepackage{algorithmic}
\usepackage{graphicx}
\usepackage{textcomp}
\usepackage{xcolor}
\usepackage{subcaption}
\usepackage{url}
\usepackage{footmisc}
\usepackage[ruled,vlined]{algorithm2e}
\usepackage[autostyle]{csquotes}

\usepackage{tikz}
\def\checkmark{\tikz\fill[scale=0.25](0,.35) -- (.25,0) -- (1,.7) -- (.25,.15) -- cycle;}

\usepackage{svg}

\usepackage{hyperref} 

\def\BibTeX{{\rm B\kern-.05em{\sc i\kern-.025em b}\kern-.08em
    T\kern-.1667em\lower.7ex\hbox{E}\kern-.125emX}}

\begin{document}

\title{Intelligent edge-based recommender system for internet of energy applications}

\author{ Aya Sayed,
         Yassine Himeur, \IEEEmembership{Member, IEEE},
        Abdullah Alsalemi, \\
        Faycal Bensaali,  \IEEEmembership{Senior Member, IEEE} and
        Abbes Amira, \IEEEmembership{Senior Member, IEEE}%
\thanks{
This paper was made possible by National Priorities Research Program (NPRP) grant No. 10-0130-170288 from the Qatar National Research Fund (a member of Qatar Foundation). The statements made herein are solely the responsibility of the authors. (Corresponding author: Aya Sayed.)}
\thanks{
A. Sayed, Y. Himeur, A. Alsalemi, and F. Bensaali are with the Department of Electrical Engineering, Qatar University, Doha 2713, Qatar (e-mail:, {as1516645@qu.edu.qa}, {yassine.himeur@qu.edu.qa}, {aa1300250@qu.edu.qa}, {f.bensaali@qu.edu.qa}).}

\thanks{A. Alsalemi and A. Amira  are with the Institute of Artificial Intelligence, De Montfort University, Leicester, UK (e-mail:, {P2621877@my365.dmu.ac.uk}, {abbes.amira@dmu.ac.uk})}
\thanks{
A. Amira is with the Department of Computer Science, University of Sharjah, UAE (e-mail:, {aamira@sharjah.ac.ae}).}
}

\markboth{IEEE System Journal }{Sayed \MakeLowercase{\textit{et al.}}: Intelligent Edge-based Recommender System for Internet of Energy Applications}


\maketitle

\begin{abstract}
Preserving energy in households and office buildings is a significant challenge, \textcolor{black}{mainly due to the recent shortage of energy resources, the uprising of the current environmental problems, and the global lack of utilizing energy-saving technologies}. Not to mention, within some regions, COVID-19 social distancing measures have led to a temporary transfer of energy demand from commercial and urban centers to residential areas, causing an increased use and higher charges, and in turn, creating economic impacts on customers. Therefore, the marketplace could benefit from developing an internet of things (IoT) ecosystem that monitors energy consumption habits and promptly recommends action to facilitate energy efficiency. This paper aims to present the full integration of a proposed energy efficiency framework into the Home-Assistant platform using an edge-based architecture. End-users can visualize their consumption patterns as well as ambient environmental data using the Home-Assistant user interface. More notably, explainable energy-saving recommendations are delivered to end-users in the form of notifications via the mobile application to facilitate habit change. In this context, to the best of the authors' knowledge, this is the first attempt to develop and implement an energy-saving recommender system on edge devices. Thus, ensuring better privacy preservation since data are processed locally on the edge, without the need to transmit them to remote servers, as is the case with cloudlet platforms.
\end{abstract}

\begin{IEEEkeywords}
Internet of Energy, COVID-19, Energy Efficiency, Smart Plug, Sensing System, Home-Assistant, Recommender System, Big Data, Data Visualization\end{IEEEkeywords}

\section{Introduction}\label{sec1}

While studying how citizens conserve electricity, most analysts and decision leaders conceptually distinguish between mitigation (e.g., unplugging chargers) and efficiency steps \cite{starke2020beyond}. However, such a two-dimensional strategy is sub-optimal from both a philosophical and a policy viewpoint since it does not consider human variations that dictate energy-saving behavior.

Conserving energy in households and office buildings is a considerable challenge, primarily due to the lack of energy resources and the uprising of the current environmental problems. Not to mention, within some regions, coronavirus 2019 (COVID-19) social distancing measures have led to a temporary transfer of energy demand from commercial and urban centers to residential areas, which has led to increased use and higher charges, creating economic impacts on customers \cite{jiang2021impacts}. Therefore, the marketplace could benefit from developing an internet of things (IoT) ecosystem that monitors energy consumption habits and promptly recommends action to facilitate energy efficiency.

\textcolor{black}{In this context, energy efficiency-driven behavioral change is a crucial element in the road to achieving feasible energy-saving \cite{alsalemi2019role}.} Therefore, a deep understanding of consumer habits is essential to find the means to transform behavior from excessive consumption towards an energy conscious.

In fact, the wide-spread use of smart meters, which glean fine-grained and frequent energy use data in comparison with traditional electricity meters, have provided more granular analytics on usage patterns \cite{himeur2020building}. On the other hand, recent hardware platforms enabled more cost-effective yet more powerful alternatives to smart meters. Those platforms can monitor energy consumption data in addition to collecting other important contextual data, which eventually leads to a larger pool of data.
The usage of emerging technology to improve energy production in homes or buildings is becoming increasingly essential for promoting energy conservation and energy protection. A number of recent developments investigate global energy-saving potential. They are seeking to build creative energy-saving technologies focused primarily on using artificial intelligence (AI), IoT devices, big data analytics, and behavioral shift  \cite{mehmood2019review}. This leads to a recent study trend named the internet of energy (IoE).

\textcolor{black}{Given the context of IoE, to investigate how improving energy behavior influences domestic energy consumption, there is a need to identify common energy consumption practices, interests, and their related costs. These can be defined as activities done at or near the given household that involve the use of electric energy, such as cooking, bathing, using a computer, watching TV, etc., \cite{smale2017social, taranunudging}. Henceforth, the concept of micro-moment is put forth as the time-based snapshots of an end-user's behavior throughout the day, which includes operating different appliances and entering/exiting a place, micro-moments provide contextually rich information about consumption.} It is also coupled with environmental metadata such as temperature, humidity, luminosity, and occupancy \cite{alsalemi2019endorsing}.
In this light, this article presents the integration of the consumer engagement towards energy-saving behavior by means of exploiting micro-moments and mobile recommendation systems (EM)$^3$ framework\footnote[1]{http://em3.qu.edu.qa/ \label{em3website}} into the Home-Assistant platform. End-users can visualize their consumption patterns as well as ambient environmental data using the Home-Assistant user interface (HAUI). More notably, energy-saving recommendations are delivered to end-users in the form of notifications via a mobile application to facilitate habit change. \textcolor{black}{In this context, to the best of the authors' knowledge, this is one of the pioneering works for developing and implementing a domestic energy-saving solution on the edge (i.e., carrying out computations and data storage locally on computing edge platforms) in contrast to relying on a cloud-based solution.}

The remainder of this paper is organized as follows. Section \ref{sec2} reviews recent work for IoE, with particular emphasis on edge/fog implementations. Section \ref{sec3} overviews the proposed framework. Section \ref{sec4} showcases the system implementation process, where the results are reported and discussed in Section \ref{sec5}. The paper is concluded in Section \ref{sec6}.

\section{Related Work}\label{sec2}
\subsection{RSs for IoE applications}
Several works have approached the issue of reducing wasted energy in the built environment from the behavioral change perspective using knowledge-based systems (i.e., RSs). In this section, we focus on discussing the most salient RS studies in the literature proposed to promote energy sustainability in buildings and reduce total energy costs. Therefore, we principally discuss existing frameworks describing energy-saving RSs, electricity plan RSs and edge/fog-based RSs. However, it is worth noting that there are other energy RSs that promote environmental benefits, such as Tomkins et al. \cite{tomkins2018sustainability}, \textcolor{black}{ which are not included in this related work-study.}

\subsubsection{Energy-saving RSs}
the power usage of residential buildings has steadily increased over the last decade. Moreover, demand for electricity in households has heavily surged during the COVID-19 pandemic period due to lock-downs and transport restrictions established by governments. However, experts suggest that users' behavior plays a significant role in wasting energy in residential buildings \textcolor{black}{\cite{paone2018impact}}. Consequently, there is a pressing need to help end-users change their consumption habits and promote more sustainable and environmentally friendly behaviors. To that end, RSs have been proposed as an effective solution to achieve the aforementioned goals \cite{himeur2021survey}. \textcolor{black}{For instance, Starke et al. \cite{starke2015saving} introduce a Rasch-based energy RS that produces personalized and engaging energy recommendations for households' end-users.} \textcolor{black}{However, the end-users' engagement with different energy-saving behaviors or measures depends on the behavioral costs (as they engage in exerting more low-cost behaviors) or difficulty of those measures.}

In \cite{kaur2019energy}, an energy RS is developed to (i) optimize energy consumption in public buildings using a multi-agent system, and (ii) consider the maximization of consumers' comfort. While In \cite{luo2017non}, a RS is proposed to help end-users select energy-saving appliances, thus creating opportunities for reducing wasted energy on the smart grid.  This was possible with the help of an energy disaggregation technique, which enables inferring the consumption of individual devices from the main consumption and hence identifying energy greedy appliances, which might be replaced with more energy-efficient ones. 

Moving forward, in \cite{alsalemi2020achieving}, an AI-based energy-saving system is developed, which uses an efficient RS. Explicitly, personalized recommendations are generated after detecting abnormal energy consumption behaviors in collected energy data using an energy micro-moment analysis. This analysis is performed at an appliance-level with reference to occupancy patterns and appliance consumption characteristics. \textcolor{black}{ Similarly, in \cite{starke2020little}, an energy-saving RS is developed, namely Saving Aid, which helps in examining the impact of RSs in promoting energy consumption behavioral change using persuasive recommendations. Accordingly, the latter have been enabled by delivering normative messages. An empirical evaluation was then conducted using contextual data, where 207 participants have been involved by comparing persuasive norm explanations, i.e., the \enquote{Saving Score}, \enquote{Global Norm}, \enquote{Similar Norm}, \enquote{Experienced Norm}}. Moving on, in \cite{sardianos2020rehab}, a context-aware RS for supporting consumers to transform their energy consumption behaviors is proposed, called recommendations for energy habits change (REHAB-C). After collecting energy footprints, occupancy patterns, and contextual data, REHAB-C is used to either trigger an energy consumption advice, postpone or cancel it and record consumers' preferences. 

In \cite{wei2020deep}, a RS named recEnergy is presented to reduce energy consumption in commercial buildings using human-in-the-loop. Specifically, the energy optimization question has been formulated as a Markov decision process and deep reinforcement learning has been investigated for learning energy efficiency recommendations. Following, recEnergy has been employed to distribute learned energy-saving actions to consumers in a commercial building in addition to collecting their feedback to further improve the recommendations.

\textcolor{black}{On the other hand, because providing end-users with only personalized recommendations can not guarantee their acceptance, the use of explanations has appeared as a potential solution to boost the acceptance rate and increase the end-user's trust in the RS \cite{zhang2018explainable, tintarev2012evaluating}. In this regard, an explainable RS is proposed in \cite{sardianos2021emergence}, which produces explainable and persuasive energy-saving advice concerning end-users' preferences and habits. The persuasive recommendations have been emphasized using economical saving prospects and/or positive ecological impacts. Following, explanations have been produced for providing the end-users with the reasons behind any energy-saving recommendation.}

\subsubsection{Electricity plan RSs}
RSs are used in this case to aid consumers select cost-effective electricity plans. \textcolor{black}{ Generally, there are different categories of electricity plans, among them: (i) the variable rate plan, in which the cost changes month to month depending on the cost of wholesale electricity and selected supplier; (ii) flat-rate plan, which allows customers to pay the same amount each month regardless of energy usage (although the energy price can be changed monthly depending on the climate situation); (iii) indexed-electricity plan, where the price paid for electricity is linked with other variables, e.g., the going rate for natural gas; (iv) green plan, which relies on providing 100\% green power to promote the generation of green energy by offsetting the usage with renewable energy; (v) time-of-use plan, where the discounts and free electricity usage are offered based on certain times or days of the week.}

Accordingly, RSs could generate advice and suggestions in either direct or indirect ways. The first approach is the most popular in online plan RSs, where different RS tools have been already developed and commercialized (e.g., Power-to-Choose \cite{Power-to-Choose2021} Energy Made Easy \cite{EME2021}, Check24 \cite{Check24-2021} and iSelect \cite{iSelect2021}). These tools focus mainly on finding and recommending the cheaper electricity plans using a deep comparison of the costs of available electricity plans, where the cost of an electricity plan is estimated using the overall electricity consumption of a consumer along with the charge rates of the plan.

In \cite{zheng2020electricity}, the authors introduce a RS to help consumers select the best electricity plans, which is implemented in two steps (i) feature formulation, and (ii) recommendation generation. The former refers to adopting a matrix recovery with energy selection rules to infer appliance energy consumption data. Then, this data has been used as the features to represent end-users' consumption behaviors. While the latter relies on the generation of collaborative filtering (CF) based recommendations using K-nearest neighbors (KNN) and adjusted similarity. This helps in recommending personal electricity plans to end-users based on the collected characteristics. Similarly, in \cite{zhang2019bayesian}, aiming at selecting the most convenient electricity retailing plans (ERP), Zhang et al. introduce a Bayesian hybrid CF-based electricity plan RS, namely BHCF-EPRS. It is implemented in a two-stage approach embedded with model-based and memory-based CF schemes. Moreover, Bayesian inference has been deployed to recover missing features and consumers' classification. Moving forward, in \cite{luo2017social}, a personalized ERP-based RS for domestic end-users is proposed, where CF is utilized to extract the pertinent energy usage features. Following, this data along with the information of their selected ERP are saved in a user knowledge database (UKD). Next, For each consumer, the RS (i) investigates his/her energy usage data, (ii) identifies consumers having similar energy usage behaviors with him/her in the UKD, and (iii) recommends the most appropriate pricing plans.

In \cite{li2019personalized}, a personalized RS to provide consumers with smart electricity tariff recommendations is proposed. It firstly collects a set of consumers’ energy usage using an advanced submetering installation. Second, the preferences of consumers in terms of tariff plans are inferred. Finally, using the inferred preference information, a matrix factorization is introduced with a CF model to recommend the most convenient tariff plans to target consumers. Similarly, in \cite{zhang2016recommending}, a customized ERP-based RS is proposed, in which cost-effective retailing plans have been produced about the users' electricity consumption footprints. Accordingly, electricity consumption has been identified using daily load profiles (DLPs) and then affected into different classes. Following, two DLPs clustering skills, plan rank-oriented, and load feature-oriented clustering have been introduced, to generate customized recommendations to help the users in selecting the more suitable ERPs.

\subsection{Edge/fog-based RSs}
\textcolor{black}{Traditional RSs cloud servers cannot dig (carefully analyze) the information to learn the users’ demands and behavior}. Therefore, edge computing and/or fog computing are computing paradigms that enable to bring the computation and storage resources into the edge devices or the network edge servers instead of having them in the remote cloud servers \cite{belli2019toward}. This could help in providing great intelligent, personalized services, and further supporting real-time implementations. To that end, increasing attention was put recently to develop edge/fog-based RSs \cite{sun2020convergence}. 

To the best of the authors' knowledge, no energy RS has been implemented on the edge devices yet. However, \textcolor{black}{there is a limited number of RS frameworks that have been proposed} for other applications and have been embedded into the edge/fog devices. For example, in \cite{gong2020edgerec}, Gong et al. introduce EdgeRec, which is an edge-based RS that aims to achieve real-time consumer perception and real-time system feedback in online shopping. Besides, to capture consumers' preferences and improve recommendations, it models heterogeneous consumer behavior sequences and adopts a context-aware reranking with behavior attention networks. In \cite{wang2019fog}, a fog-based hybrid RS is proposed for addressing the problems of information overload. Explicitly, it can abstract pertinent data from the fog environment and further provide the users with localized suggestions for improving the system performance. In \cite{ibrahim2020fog}, aiming to help the users to discover the most appropriate materials to enhance their performance in an e-learning system, a fog-based RS is proposed. In this regard, the improvement is ensured by achieving a fast response time and increasing the security to overcome the issues related to the personalization and synonymy in e-learning systems.

\subsection{Discussion}
Although the development of RSs is receiving an increasing interest nowadays in the building energy sector, different issues are still open and need to be addressed to develop reliable, engaging, and real-time RSs. For instance, most existing energy RS frameworks are implemented on cloud servers, but the growing use of cloud computing and networking results in high bandwidth consumption, leading to higher energy usage \textcolor{black}{\cite{hannan2018review}}. Moreover, the bandwidth constraints, Internet dependency, and latency issues due to data transmission to cloud servers significantly hinder the design of real-time RSs using cloud computing \textcolor{black}{\cite{cao2020overview}}. On the other hand, the transmission and storage of sensitive personal data on the cloud raise additional security and privacy problems, which have not been addressed in most existing frameworks \textcolor{black}{\cite{liang2021mobile}}. To fill that gap, we propose in this paper a novel energy-saving framework based on an explainable RS, which is implemented on the edge. It includes different components, which are (i) the smart plug used to collect data, (ii) the anomaly detection module, (iii) the explainable recommendation engine, and (iv) the end-user app used to receive recommendations and visualize contextual data. Thus, the proposed framework presents better privacy preservation potential than existing RSs due to processing data on edge devices close to the end-user without the need to transmit them to remote servers. Moreover, this reduces the latency as data are processed locally and supports real-time or near real-time implementations. \textcolor{black}{On the other hand, the persuasive recommendations in the RS, that emphasize economic benefits are consistent with the one used by Starke et al. \cite{starke2015saving}.}


\section{Framework Overview} \label{sec3}
\begin{figure}[!t]
\centering
\includegraphics[width=0.85\columnwidth]{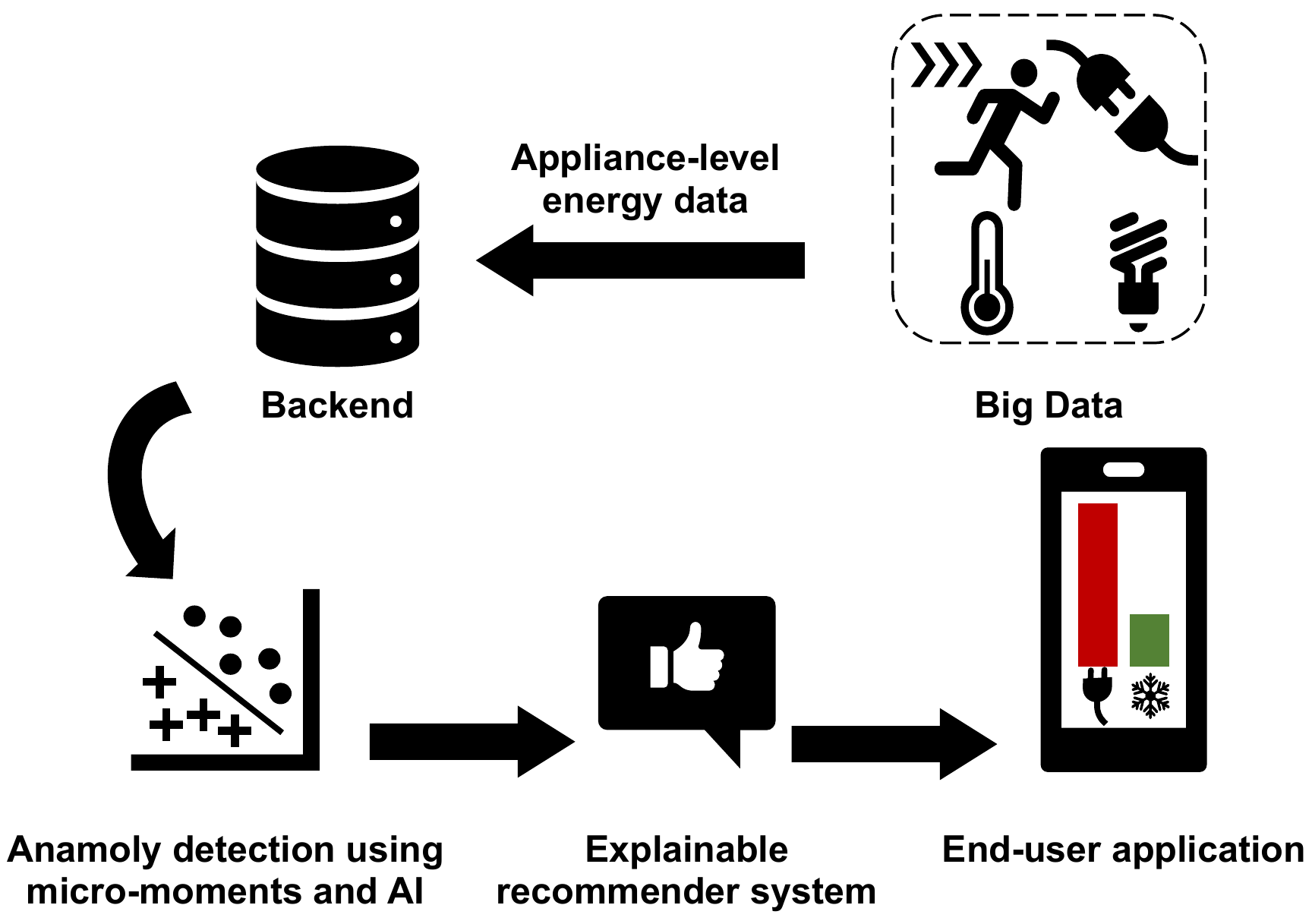}
\caption{{Overview of the (EM)$^3$ framework.}}
\label{em3-overview} 
\end{figure}

The (EM)$^3$ framework is designed to advance the state-of-the-art evidence-based, technology-enabled energy efficiency recommendation systems to enhance domestic energy efficiency. As a result, the (EM)$^3$ framework comprises a variety of sub-systems that work together to accomplish this goal \cite{alsalemi2019endorsing}. The data flow between the sub-systems is shown in Fig. \ref{em3-overview}. Several sensing instruments (e.g., smart plugs, temperatures, humidity, motion sensors, etc.) are used to collect different types of data transmitted wirelessly to the back-end, where they will be classified into a group of micro-moments. Classified micro-moments are used as inputs to the RS.

\subsection{System back-end and data collection}
This sub-system primarily back-ends composite data from multiple sensors, and additionally keeps processed information on consumer energy profiles, micro-moments, and recommendations. Sensors (e.g., temperature, humidity, illumination, etc.) report user presence and environmental conditions both within and outside the tracked site, while smart meters record electric power usage per unit, resulting in a stream of measurement data collected in the platform database. This is achieved by utilizing an open-source automation platform called Home-Assistant\footnote[2]{https://www.home-assistant.io/ \label{home-assistant}}. The platform prioritizes privacy and local control. The host for running Home-Assistant, in this case, is a Raspberry Pi 4 Model B, which also acts as the server in the system architecture. Home-Assistant uses message queuing telemetry transport (MQTT) to connect with other devices. To elaborate, MQTT is a lightweight messaging protocol used to exchange messages among devices.

\subsection{Proposed RS}
\begin{figure}[]
    \centering
    \includegraphics[width=8.5cm,height=7.5cm]{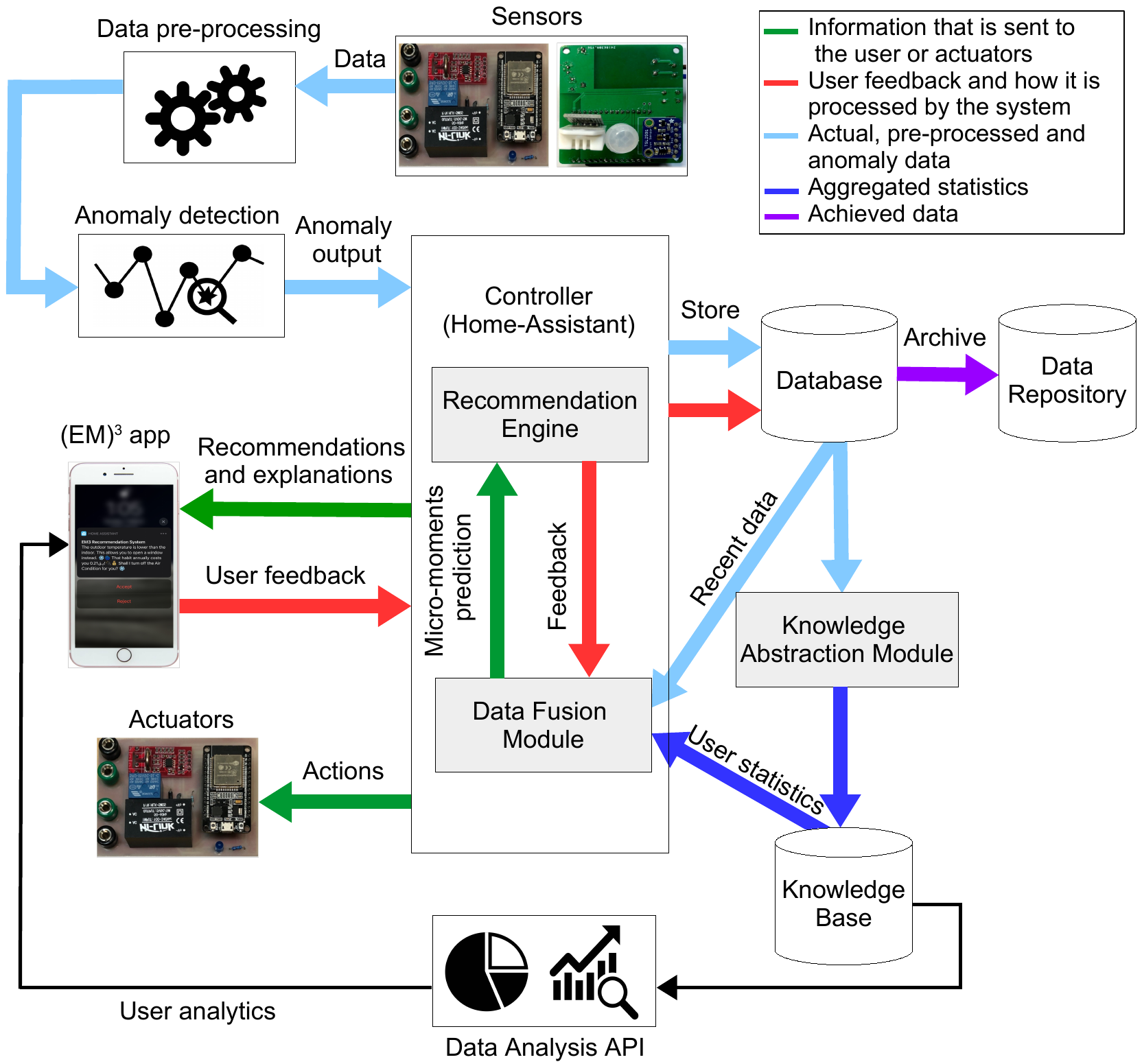}
    \caption{The core architecture of the RS with the explainable recommendation extensions.}
    \label{fig:recommende-system}
\end{figure}

The ecosystem depicted in Fig. \ref{fig:recommende-system} serves as the foundation for the system's explainable recommendations, consisting of sensors, smart meters, actuators, and orchestrating software that works together to promote energy conservation through smart on-time recommendations based on the habit loop theory of behavioral change. Data collection and storage are detailed in the previous sub-section. 

An additional knowledge abstraction module (KAM) has been built to effectively manage the large volume of data produced by the sensors and smart meters. It is made up of scripts that process sensor data streams in real-time and detect micro-moments. When the user leaves the room, or when the external conditions meet the user expectations, or when a device has been used frequently, and so on. Thus, micro-moments are defined as moments of particular importance to the user. The system's knowledge base stores the observed micro-moments, as well as information about the user's preferences. The system's explainable recommendations architecture is assisted by an adequate data model, which organizes real-time data gathered by sensors, aggregated data that summarize recent application use, and room presence data (for a few weeks) at 5-min granularity. In the former, the comprehensive gathered data is stored in the dataset for a few months, while the latter is regularly updated to represent updated usage preferences at a particular moment and stored in what is referred to as the knowledge base. For archiving purposes, older sensor data is transferred into a Data Repository.

To generate recommendations, the action triggering module (ATM) receives real-time room occupancy, appliance consumption, and environmental details, as well as information on user behaviors. The ATM is written as scripts in Home-Assistant. Custom scripting in Python and other languages is supported, and the software exposes various APIs for interacting with third-party applications and frameworks. The main medium for delivering recommendations to end-users is their mobile and Home-Assistant application, which presents recommendations and gathers user feedback (i.e., accept, reject, or ignore). 

Initially, when the recommendation triggers, the system in the background collects user preferences from the knowledge base and reviews recent sensor entries in the database to detect whether a recommendation is at the appropriate micro-moment. Using the Home-Assistant mobile application, produced recommendations are displayed on the user's smartphone. The suggestions obtained are followed by explanations and more compelling facts about an action's impact.

\begin{figure}[!t]
\centering
\begin{subfigure}{.15\textwidth}
  \centering
  \includegraphics[width=0.85\linewidth]{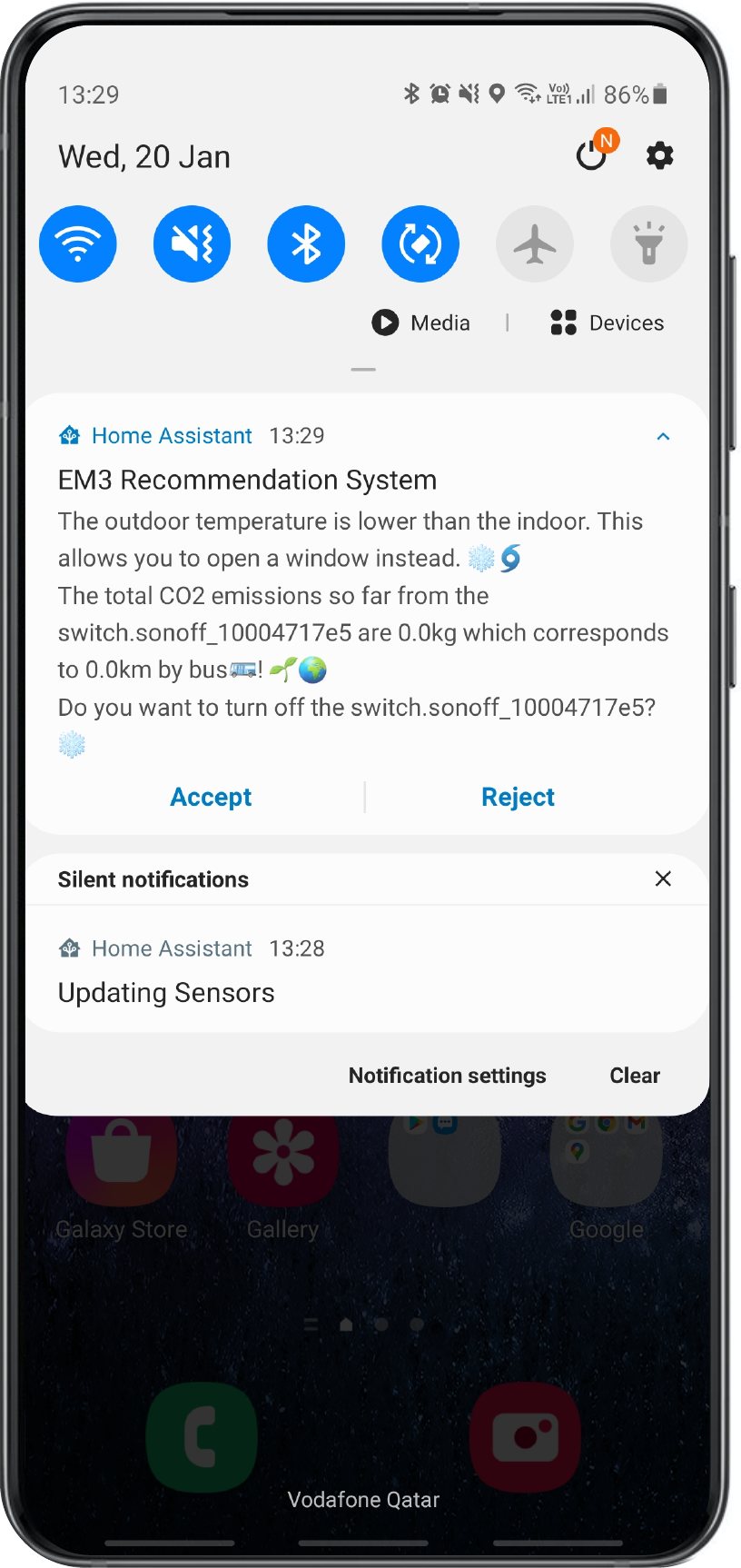} 
  \caption{}
  \label{fig:sub-first-em3app}
\end{subfigure}
\begin{subfigure}{.15\textwidth}
  \centering
  \includegraphics[width=0.85\linewidth]{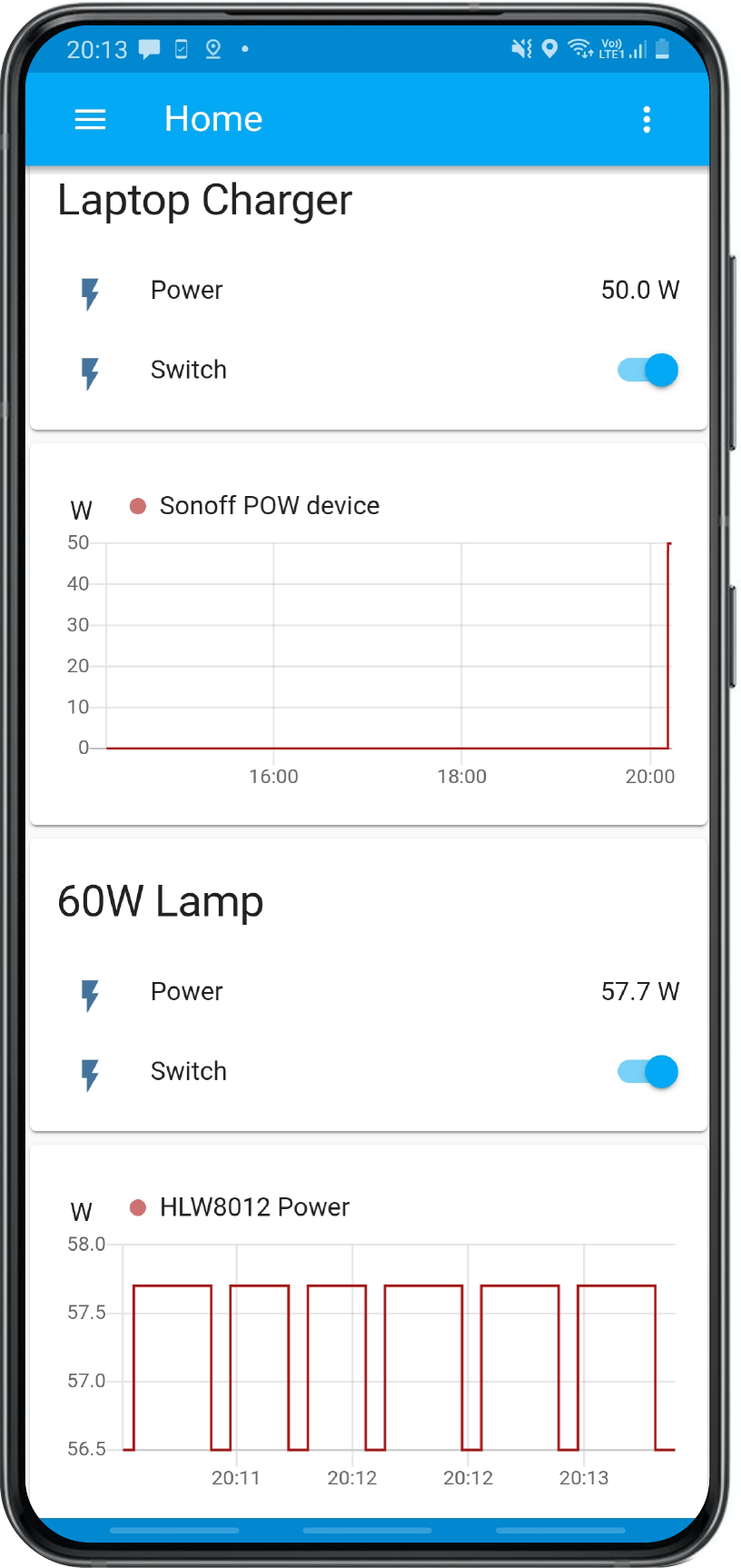} 
  \caption{}
  \label{fig:sub-second-em3app}
\end{subfigure}
\begin{subfigure}{.15\textwidth}
  \centering
  \includegraphics[width=0.85\linewidth]{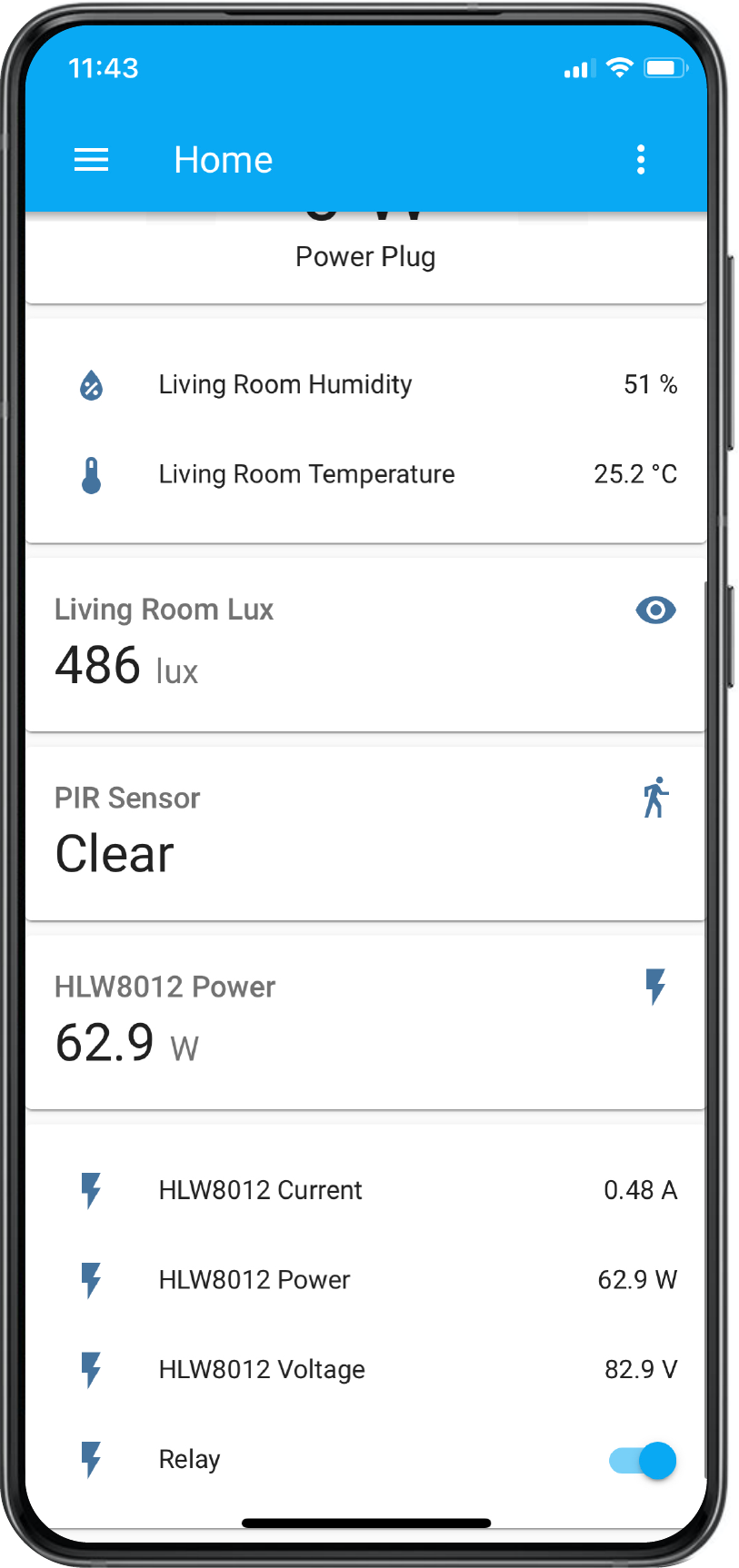} 
  \caption{}
  \label{fig:sub-third-em3app}
\end{subfigure}
\caption{Home-Assistant mobile UI, denoting the (a) generated recommendations, (b) power consumption visualization with appliance control, and (c) environmental data.}
\label{em3-appliaction}
\end{figure}

\subsection{Data analytics and visualization}
Appropriate data visualization of energy usage is essential in teaching end-users about their energy utilization habits and suggesting areas of change. The mobile application supported by the Home-Assistant platform collects information and analysis from the back-end and shows condensed information on energy usage and weather data accordingly. Further, the mobile application presents the suggestions produced by the RS and enables simple remote control of the appliance. \textcolor{black}{Fig. \ref{em3-appliaction}, demonstrates the mobile UI developed for monitoring energy consumption and environmental conditions, designed as scripts in the open-source home-automation platform of Home-Assistant. This platform uses custom scripts in Python (and other languages) and enables diverse APIs to communicate with external systems and applications. The key channel for providing the end-user with the recommendations is his/her smartphone and mobile UI. The latter serves for displaying recommendations and collecting end-user responses (i.e., reject, accept, or ignore). Put simply, by using this mobile app, the end-user can receive notifications regarding their energy consumption with areas of improvement. Besides, end-users can visualize their consumption, along with the surrounding environmental data.}

Real-time room occupancy, appliance consumption, and environment-related data along with knowledge about user habits are fed to the Action Triggering Module (ATM). The ATM is developed as scripts in the open-source home automation platform of Home-Assistant. The platform allows custom scripting in Python and other languages and exposes several APIs for communicating with external applications and systems. The key channel for communicating recommendations to the end-user is his/her smartphone and the Telegram application, which displays recommendations and collects user responses (i.e., accept, reject or ignore).

\begin{figure}[!t]
\centering
\includegraphics[width=0.49\linewidth]{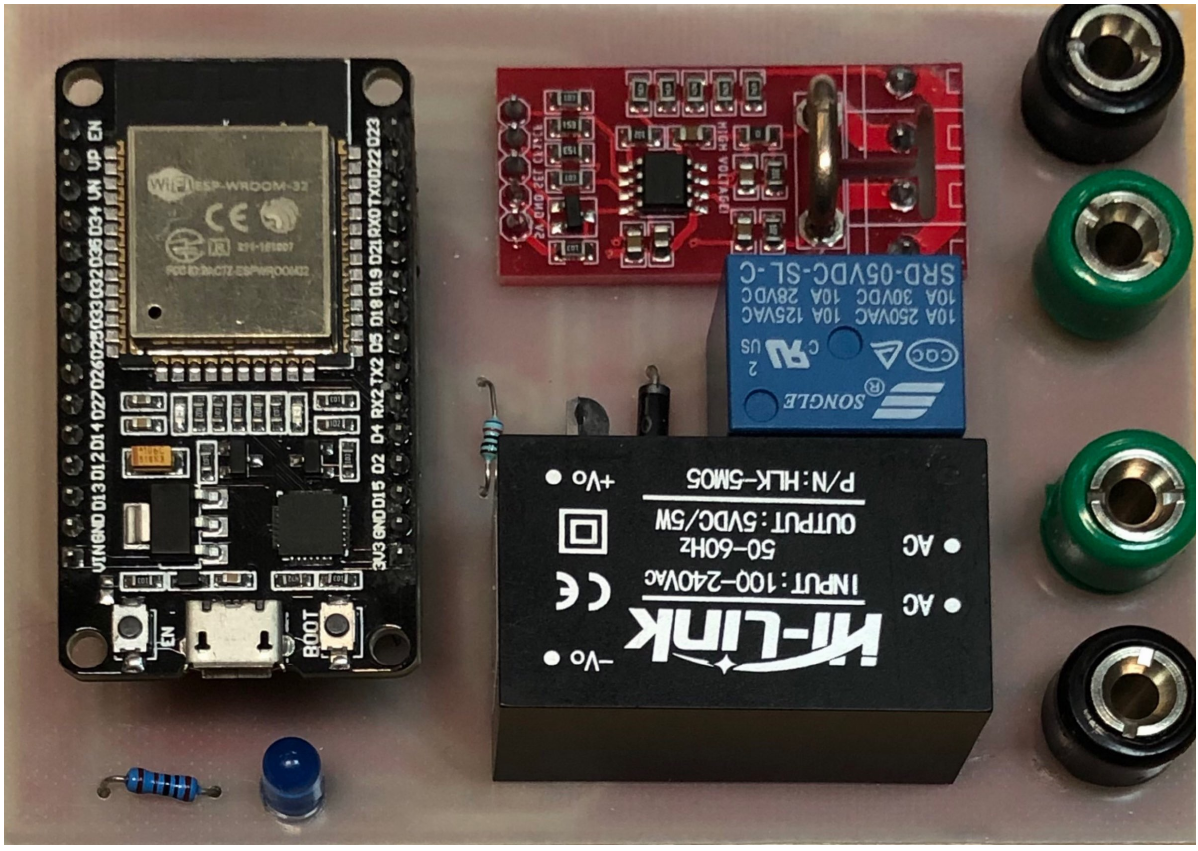} 
\includegraphics[width=0.49\linewidth]{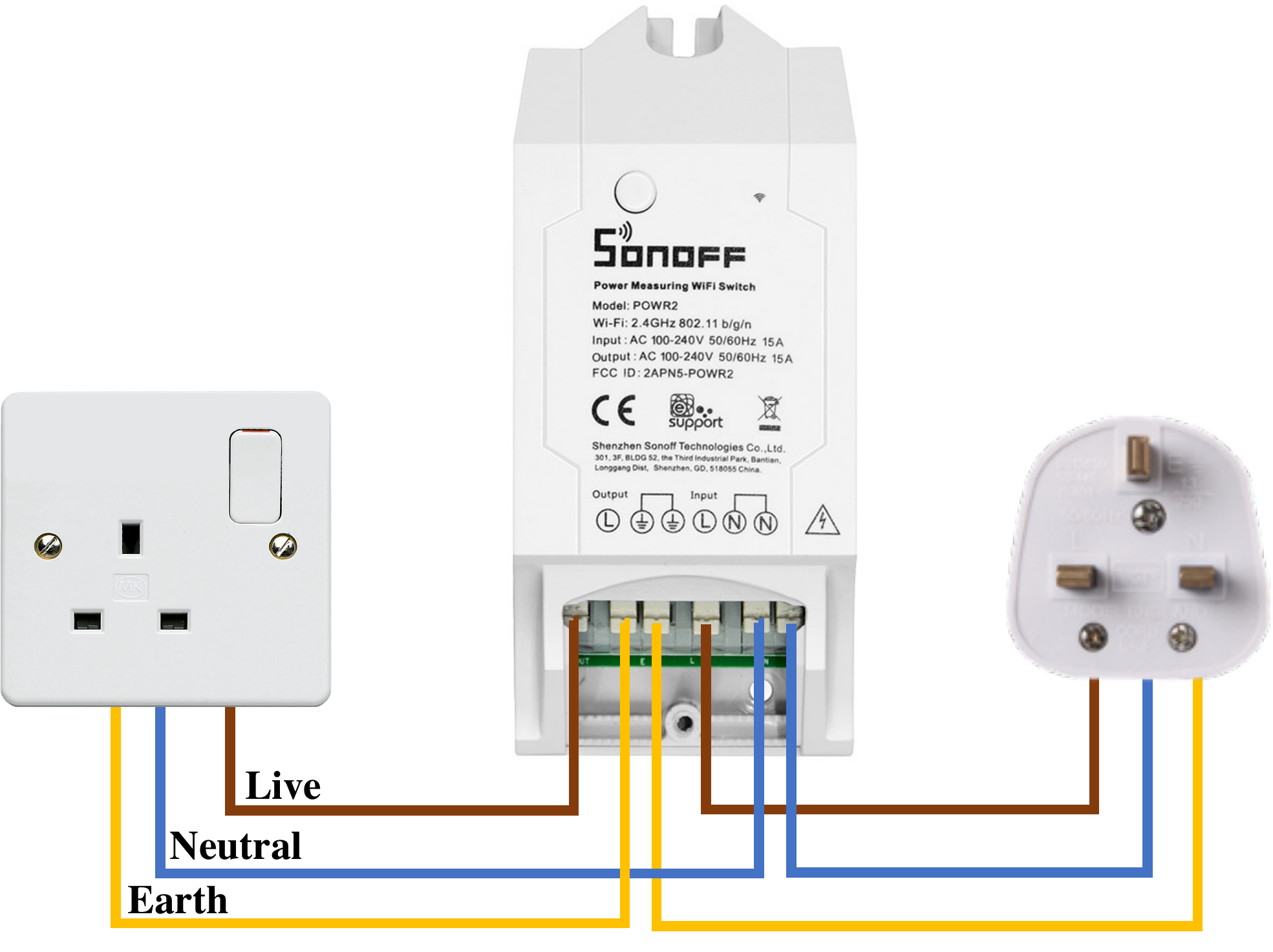}
\caption{Smart plug alternatives, including, the (left) (EM)$^3$ smart plug, and (right) sonoff POW device.}
\label{smart-plug}
\end{figure}

\section{Implementation} \label{sec4}
The overall aim of the (EM)$^3$ framework is to enhance end-users' energy consumption behaviors, and it does so by, firstly, providing guidelines for energy-saving practices sent to the end-users' mobile device at the appropriate time and secondly, providing energy usage analytics that demonstrates to the consumer the advantages of improving their behaviors. The aforementioned processes are enabled by data sources originating from the sensors allocated in the power and environmental modules. \textcolor{black}{Provided that the following components have been evaluated individually, within the overall framework and yielded acceptable performance, hence were selected to take-part of the final framework. The framework was additionally assessed by creating a test-bed shown in Fig. 7, with the primary goal of measuring communication latency and comparing the efficiency of smart plug alternatives to a reliable watt-meter. In [39], the remaining sensors were calibrated and evaluated by utilizing a standard measuring instrument. The list of components used in the hardware implementation is found in Table. \ref{components-table}. In Section V a detailed examination of the components can be found.}

\begin{table}[!t]
\centering
\caption{List of sensing and actuation components.}
\label{components-table}
\begin{tabular}{l@{\hskip 0.1in} l}
\hline
Component Name & Operating Range                                                                            \\ \hline
DHT22          & Temperature: -40$^{\circ}$C-80$^{\circ}$C and Humidity: 0-100\%                                \\
TSL2591        & 188${\mu}$Lux-88,000Lux                                                                         \\
AM312          & 3-5 meters                                                                                       \\
HLW8012        & \begin{tabular}[c]{@{}l@{}} Maximum power: 3500W\end{tabular} \\
5V Relay       & 10A and 250VAC                                                                                     \\
Sonoff POW     & \begin{tabular}[c]{@{}l@{}} Maximum power: 3500W\end{tabular}   \\ \hline
\end{tabular}
\end{table}

\subsection{Edge IoE smart plug }
A proper sensing system is needed to compute the power consumption of a given appliance. Two power measuring alternates are employed: the HLW8012 power sensor and the Sonoff POW device. Both options are equipped with a remote control mechanism to enable users to manage the status of their appliances. The data recovered from the smart plugs are transmitted to an edge device to analyse the data acquired and output an energy-saving recommendation. An illustration of the smart plugs used is found in Fig. \ref{smart-plug}.

\subsection{Environmental module}
\begin{figure}[!t]
\centering
  \includegraphics[width=0.32\linewidth]{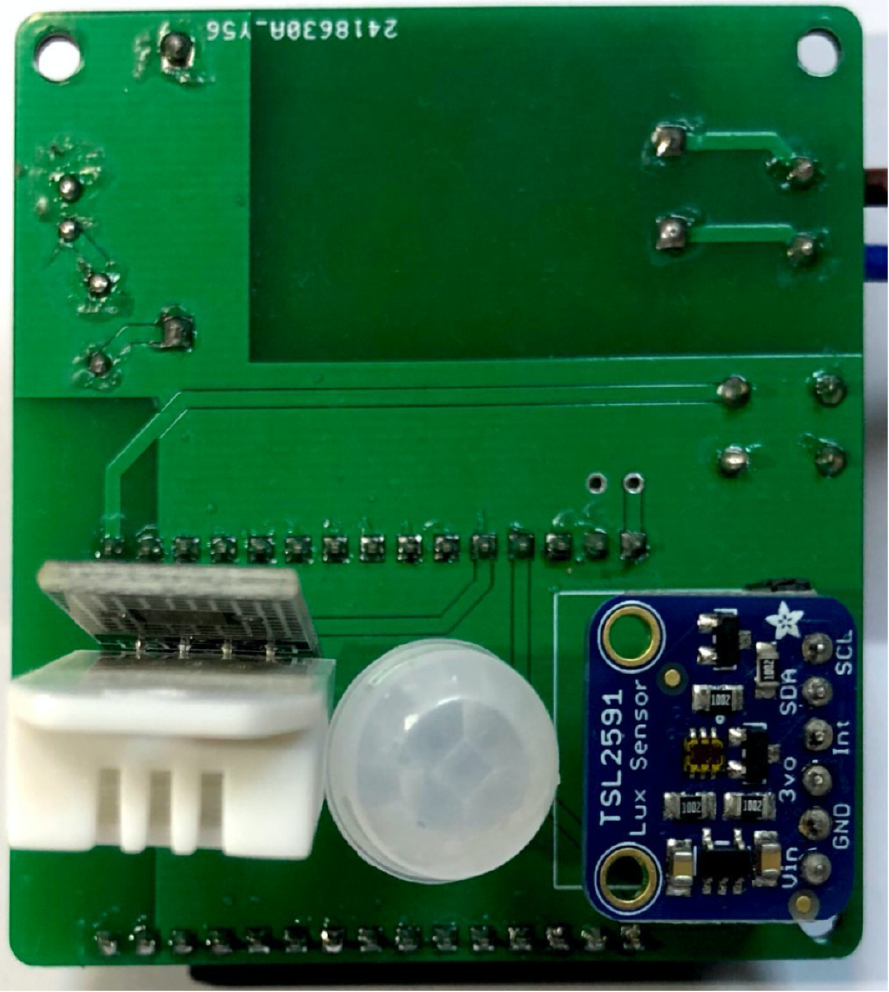}  
  \includegraphics[width=0.32\linewidth]{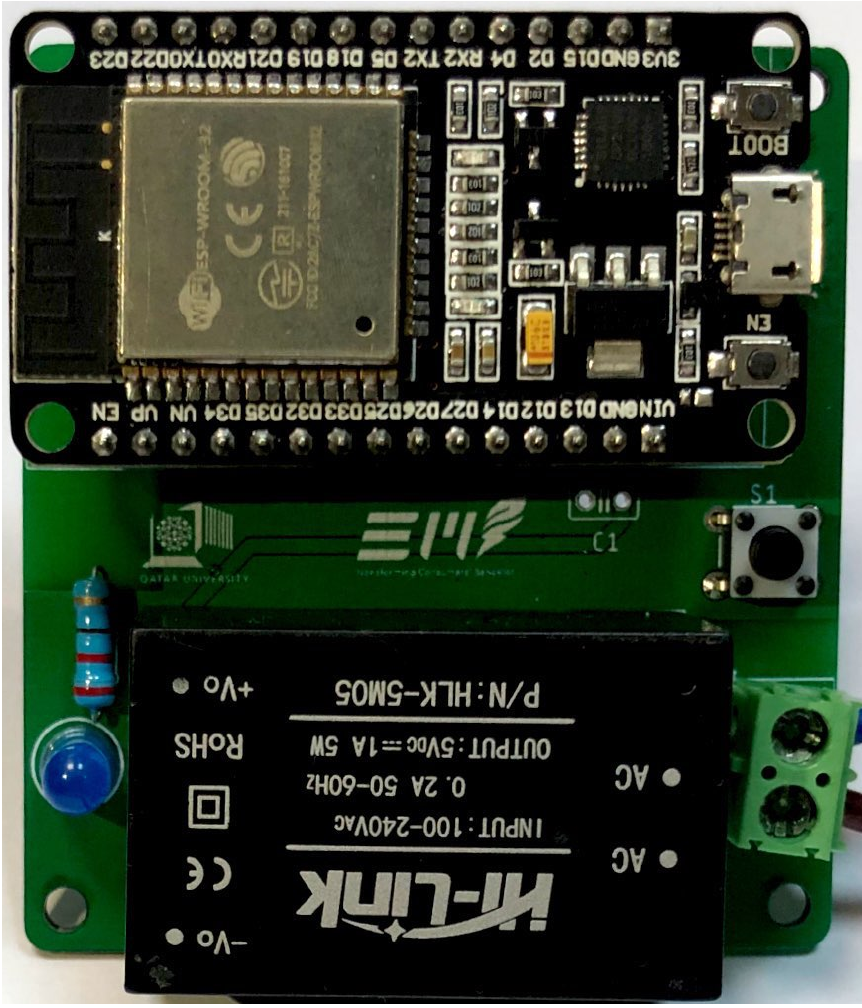}  
\caption{Environmental module, with, the (left) front view, and (right) back view.}
\label{envo-module}
\end{figure}

Collecting contextual information can be of great use to create more meaningful recommendations. Hence, temperature, humidity, luminosity, and occupancy of a given household space are all measured using the environmental module. The data is forwarded for further analysis in real-time to the back-end. To elaborate on the significance of gathering such data, determining whether the end-user occupies a space is essential for power usage management. This knowledge will facilitate recognizing the end-user's habits and, as a result, enable more educated suggestions. The environmental module is shown in Fig. \ref{envo-module}.

\subsection{Data analysis process}
\subsubsection{Feature extraction}
To define abnormal consumption, a micro-moment analysis is conducted on the collected power consumption data. Therefore, five classes are identified using Algo. \ref{algo1} to describe power consumption observations of each appliance, they are defined as: \enquote{class 0: good usage, which refers to the case when power consumption is $<95\%$ maximum of active consumption rate}; \enquote{class 1: turn on the appliance}; \enquote{class 2: turn off the appliance}; \enquote{class 3: excessive consumption}, which stands to the case of power consumption that is $>95\%$ maximum of active consumption rate; and \enquote{class 4: consumption while outside}, this is limited to detecting abnormal consumption of an ensemble of device categories, including the air conditioner (A/C), television, light lamp, desktop/laptop, and fan, where the presence of the end-user during their operation is a must to not be considered as abnormal. 

The micro-moment features M$^{2}$F are extracted based on analyzing the occupancy profile ($O$) and power consumption ($p$) of each device in reference to device active consumption range ($DACR$), device operation time ($DOT$), and device standby power consumption ($DSPC$). 
Then, the appliance operation parameters are called, including $DACR$, $DOT$ and $DSPC$. Table \ref{AppSpecifications} presents an example of different appliance parameter specifications used in the rule-based algorithm to extract power consumption micro-moments. 

\begin{algorithm}[htbp]
\SetAlgoLined
\KwResult{M$^{2}$F: micro-moment feature vector}
Read $p, O, DACR, DOT$, , $DSPC$ and $O_{T}$: operation time; \\
Initialization: M$^{2}$F = $\emptyset$
\While{$t \leq l$ (\textnormal{with} $l$ \textnormal{is the length of the power signal})}{
\textbf{Rule 1:} Non-excessive usage \\
\textbf{if} $p(t) \geq min(DACR) $ \textbf{and} $ p(t) \leq 95\% \times$ max$(DACR)$ \\
\hspace*{10mm} M$^{2}$F(t) = 0 (Good usage); \\
\textbf{Rule 2:} Switching on a device \\
\textbf{if} $p(t) \geq$ min$(DACR)$ \textbf{and}  $p(t-1) \leq$ max$(DSPC)$ \\
\hspace*{10mm} M$^{2}$F(t) = 1 (Turn on device); \\
\textbf{Rule 3:} Switching off a device \\
\textbf{if} $p(t) \leq$ max $(DSPC)$ \textbf{and} $p(t-1) \geq$ min $(DACR)$  \\
\hspace*{10mm} M$^{2}$F(t) = 2 (Turn off device);\\
\textbf{Rule 4:} Consumption exceeds 95\% of DACR or DOT \\
\textbf{if} $p(t) \geq 95\% \times$ max$(DACR)$ \textbf{or} $O_{T}(t) \geq DOT$   \\
\hspace*{10mm} M$^{2}$F(t) = 3 (Excessive consumption);   \\
\textbf{Rule 5:} Consumption without presence of the end-user \\
\textbf{if} $O(t) = 0$ \textbf{and} $p(t) \geq 0.95 \times$ $DSPC$    \\
\hspace*{10mm} M$^{2}$F(t) = 4 (consumption while outside); \\
}
\caption{Proposed rule-based algorithm for extracting power consumption micro-moment features.}
\label{algo1}
\end{algorithm}

\begin{table} [t!]
\caption{Power consumption specifications for different home appliances.}
\label{AppSpecifications}
\begin{center}

\begin{tabular}{lccc}
\hline
{\scriptsize Appliance \ \ \ \ \ \ } & {\scriptsize DOT } & {\scriptsize \
DACR (watts)} & {\scriptsize DSPC (watts) } \\ \hline
{\scriptsize Air conditionner } & {\scriptsize 15 h 30 min} & {\scriptsize %
1000} & {\scriptsize 4} \\ 
{\scriptsize Microwave} & {\scriptsize 1h} & {\scriptsize 1200} & 
{\scriptsize 7} \\ 
{\scriptsize Oven} & {\scriptsize 3h} & {\scriptsize 2400} & {\scriptsize 6}
\\ 
{\scriptsize Dishwasher} & {\scriptsize 1h 45 min} & {\scriptsize 1800} & 
{\scriptsize 3} \\ 
{\scriptsize Laptop} & {\scriptsize 12 h 42 min \ } & {\scriptsize 100} & 
{\scriptsize 20} \\ 
{\scriptsize Washing machine \ \ \ \ \ \ \ } & {\scriptsize 1h} & 
{\scriptsize 500} & {\scriptsize 6} \\ 
{\scriptsize Light} & {\scriptsize 8 h} & {\scriptsize 60} & {\scriptsize 0}
\\ 
{\scriptsize Television} & {\scriptsize 12 h 42 min} & {\scriptsize 65} & 
{\scriptsize 6} \\ 
{\scriptsize Refrigerator} & {\scriptsize 17 h 30 min} & {\scriptsize 180} & 
{\scriptsize 0} \\ 
{\scriptsize Desktop} & {\scriptsize 12 h 42 min} & {\scriptsize 250} & 
{\scriptsize 12} \\ \hline
\end{tabular}

\end{center}
\end{table}

Following, the recommendations that aim to steadily modify the user's routine, for instance, decreasing the energy footprint of a repetitive action, are produced. To do so, following the outcome of the anomaly detection, the suggestions must be delivered at the right moment of time. The RS seeks to optimize the user incentive smoothly, so it focuses on energy-consuming routines that occur infrequently or routines that occur regularly, which offers a significant accumulated incentive for a minor behavioral shift.

The suggestions in (EM)$^3$ are founded on the rules derived from an overview of user behavior in the past. In short, user activities and their circumstances, as registered in user activity records, are abstracted and analyzed in accordance with the rules of association rule learning. The rule is based on a series of user actions $(u_a)$ and circumstances $(c)$ (e.g., time day, temperature, user status, etc.) that are regularly replicated in user logs. The right-hand side of the rule (RHS) involves constantly replicated user behavior (e.g., the user switched on or off the device, extensively used device, etc.) and the left-hand side (LHS) refers to the conditions that hold when this action occurs. To sum up, the produced recommendations are described as being: (i) customized since they target the habitual activities of the user, (ii) presented in the proper way (i.e., at the right time where relevant circumstances take place), and, (iii) used to serve a particular purpose and that is to significantly increase energy efficiency.

\subsubsection{Recommendations’ delivery}
The main medium for delivering recommendations to end-users is their mobile and Home-Assistant application, which presents recommendations and gathers user feedback (i.e., accept, reject, or ignore). 

\subsubsection{Recommendations’ acceptance}
\begin{figure}[!t]
\centering
\includegraphics[width=7.5cm, height=5cm]{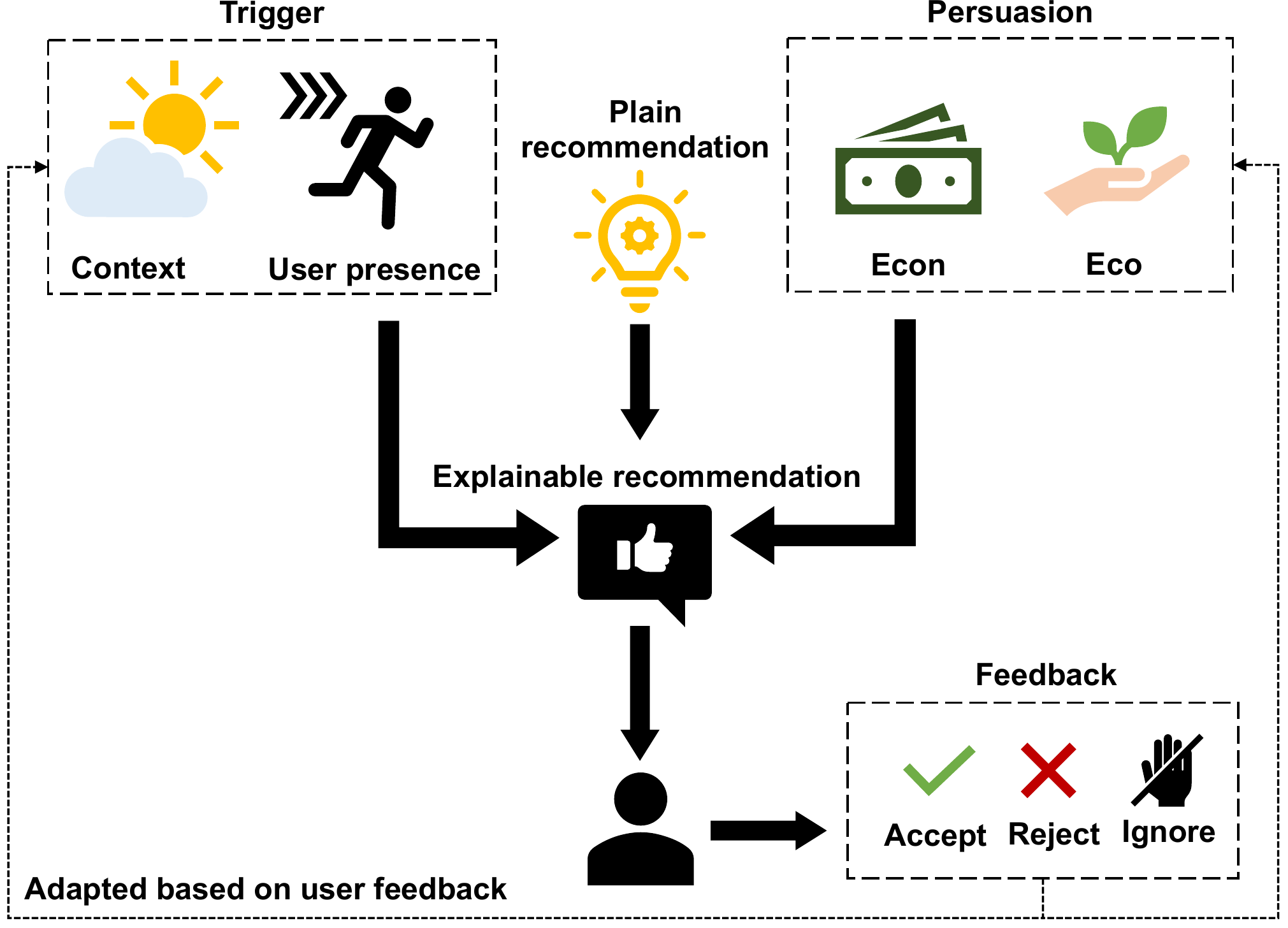}
\caption{{The flow of explainable recommendations. }}
\label{explainable-recommendation} 
\end{figure}

Aside from the recommendation's purpose, a number of details that educate the consumer about the advantages of taking a particular action will help to enhance the recommendation's acceptance. A compelling fact supports the advice and aids the consumer in creating a more energy-sustainable profile. The RS flow to produce explainable recommendations is shown in Fig. \ref{explainable-recommendation}. \textcolor{black}{As observed from the figure, the suggestions are customized by considering the consumers’ feedback before generating future recommendations.}

Since providing additional convincing information to the user will improve the user's adoption of advice, the following facts are included: which are the (i) eco (ecological) style of persuasion facts which are based on the energy consumption's environmental effect, and, (ii) econ (economical) style of persuasion facts which are based on the energy consumption's economical effect (saving cost).

A portion of the recommendation message is devoted to explaining the rationale behind the triggering of each recommendation. Two main reasons are responsible for triggering a recommendation, including:
\begin{itemize}
    \item {Recommendations are fired because of the awareness that the user has left the room and kept an appliance on, thereby wasting electricity without need, and}
    \item{Recommendations are fired when the user is present and has an appliance on in the room, although the outdoor conditions (i.e., light or temperature) make it possible to prevent unnecessary use, for instance, through opening a window to cool the space or allowing natural light.}
\end{itemize}

\section{Results and Discussion} \label{sec5}
This section is dedicated to evaluating the system's performance in four key areas, such as smart plug accuracy, anomaly detection, and RS scenarios, as well as comparing the overall system outcomes to similar existing works.

\begin{figure}[!t]
\centering
\includegraphics[width=8.5cm, height=6cm]{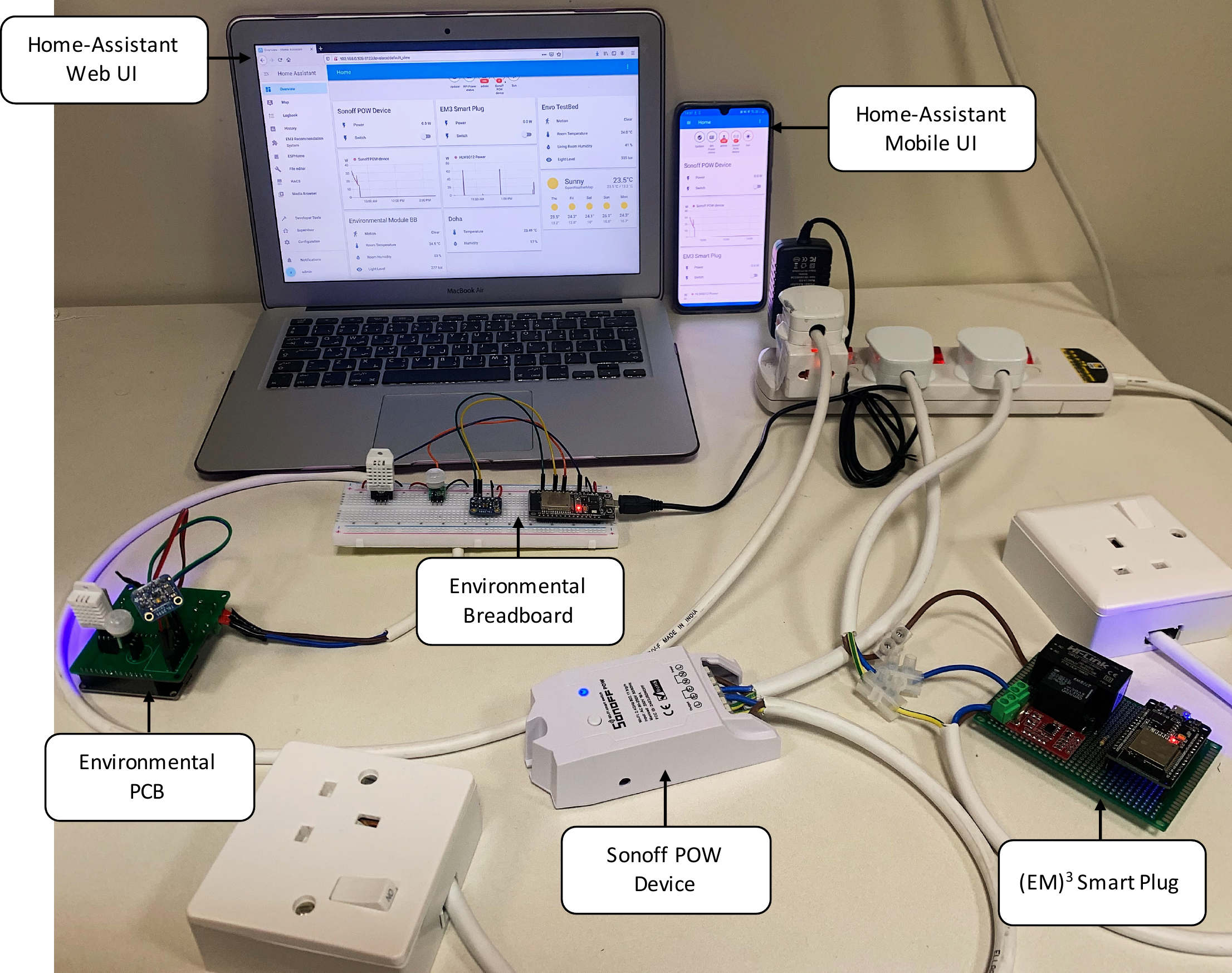}
\caption{{Overview of the test-bed. }}
\label{test-bed} 
\end{figure}

\subsection{Test bench overview}
Fig. \ref{test-bed}, shows the layout of the test-bed. It comprises a web or a mobile application UI, an edge server with Home-Assistant running, and the different sensors capturing either spatial data or power usage. The test-bed was created with the primary goal of measuring communication latency and comparing the efficiency of smart plug alternatives to a reliable watt-meter. The remaining sensors were compared to a standard measuring device in \cite{alsalemi2019boosting}. 

\subsection{Comparative analysis between smart plugs}
Using the PX110 watt-meter\footnote[3]{http://www.farnell.com/datasheets/3649.pdf \label{px110}}, to test the accuracy of the smart plugs alternatives. The meter has an accuracy of ±1.5\% with real power. Table. \ref{accuracy-results}, illustrates the accuracy of the smart plug alternatives attached to various home appliances, as well as the reference used. The test was given a time frame of 45-60 seconds, depending on the appliance.

\begin{table}[!t]
\caption{Smart plugs accuracy results.}
\label{accuracy-results}

\begin{tabular}{lll}
\hline
Appliance &   (EM)$^3$ Smart Plug Accuracy & Sonoff POW Accuracy \\\hline
{60W lamp} & \multicolumn{1}{c}{99.28\%} & \multicolumn{1}{c}{97.58\%}\\
{Air cooler} & \multicolumn{1}{c}{98.87\%} & \multicolumn{1}{c}{96.51\%}\\
{Monitor} & \multicolumn{1}{c}{96.55\%} & \multicolumn{1}{c}{95.95\%} \\ 
{Kettle} & \multicolumn{1}{c}{99.12\%} & \multicolumn{1}{c}{96.93\%} \\ 
{Laptop} & \multicolumn{1}{c}{96.96\%} & \multicolumn{1}{c}{97.44\%} \\ 
{Overall} & \multicolumn{1}{c}{98.16\%} & \multicolumn{1}{c}{96.88\%} \\ \hline
\end{tabular}

\end{table}

\subsection{Anomaly detection scenarios}
To validate and automate the anomaly detection process, we proceed with the tests on (i) real-world data collected via the QUD (Qatar University dataset) and DRED (Dutch residential energy dataset) campaigns \cite{himeur2020building}, and (ii) simulated dataset (SiD) \cite{alsalemi2019endorsing}. \textcolor{black}{QUD was gathered at Qatar University energy lab by recording power consumption footprints and other indoor climate conditions, including occupancy, temperature, and humidity at a sampling interval ranging from 3 sec to 30 min. DRED includes energy consumption, occupancy, and environmental data of a typical household in the Netherlands. Electricity footprints were recorded from 12 distinct devices for both aggregated- and appliance-levels, at the sampling rates of 1 min and 1 sec, respectively. SiD includes power usage patterns of 2 years, which were generated using hourly load profiles at an appliance-level. The manufacturer appliances' specifications were utilized for defining the energy consumption footprints of every device in watts. Occupancy data were generated as well using diverse occupancy rules that represented the percentage of end-users' presence per day in a typical house.}

Moreover, the empirical evaluation is conducted in comparison with various machine learning classifiers (i.e., Logistic regression (LR), linear discriminant analysis (LDA), support vector machine (SVM), naive Bayes (NB) decision tree (DT), random forest (RF), multi-layer perceptron (MLP), k-nearest neighbors (KNN), and deep neural networks (DNN)). \textcolor{black}{Table \ref{parameter-set} summarizes the parameter settings used in the assessment process in order to ensure the best performance for each classifier.}

\begin{table*}[!t]
\caption{\textcolor{black}{Parameter settings used in the assessment. }}
\label{parameter-set}

{\color{black}
\begin{tabular}{lllllllll}
\hline
& {\small LDA \ \ \ \ \ } & {\small PCA \ \ \ \ \ } & {\small RF} & {\small %
DT} & {\small DNN} & {\small SVM} & {\small KNN} & {\small EBT} \\ 
\cline{2-9}
{\small Parameter} & {\small N/A} & {\small Nonlinear} & {\small Gaussian }
& {\small Fine, 100} & {\small 30 hidden } & {\small Gaussian} & {\small %
K\thinspace =\thinspace 1, Euclidean } & {\small Bagged, 30 learners} \\ 
{\small setting} &  & {\small kernel} & {\small classifier} & {\small splits}
& {\small layers} & {\small kernel} & {\small distance} & {\small 42 k splits%
} \\ \hline
\end{tabular}
}
\end{table*}

\begin{figure}[t!]
\begin{center}
\includegraphics[width=0.45\textwidth]{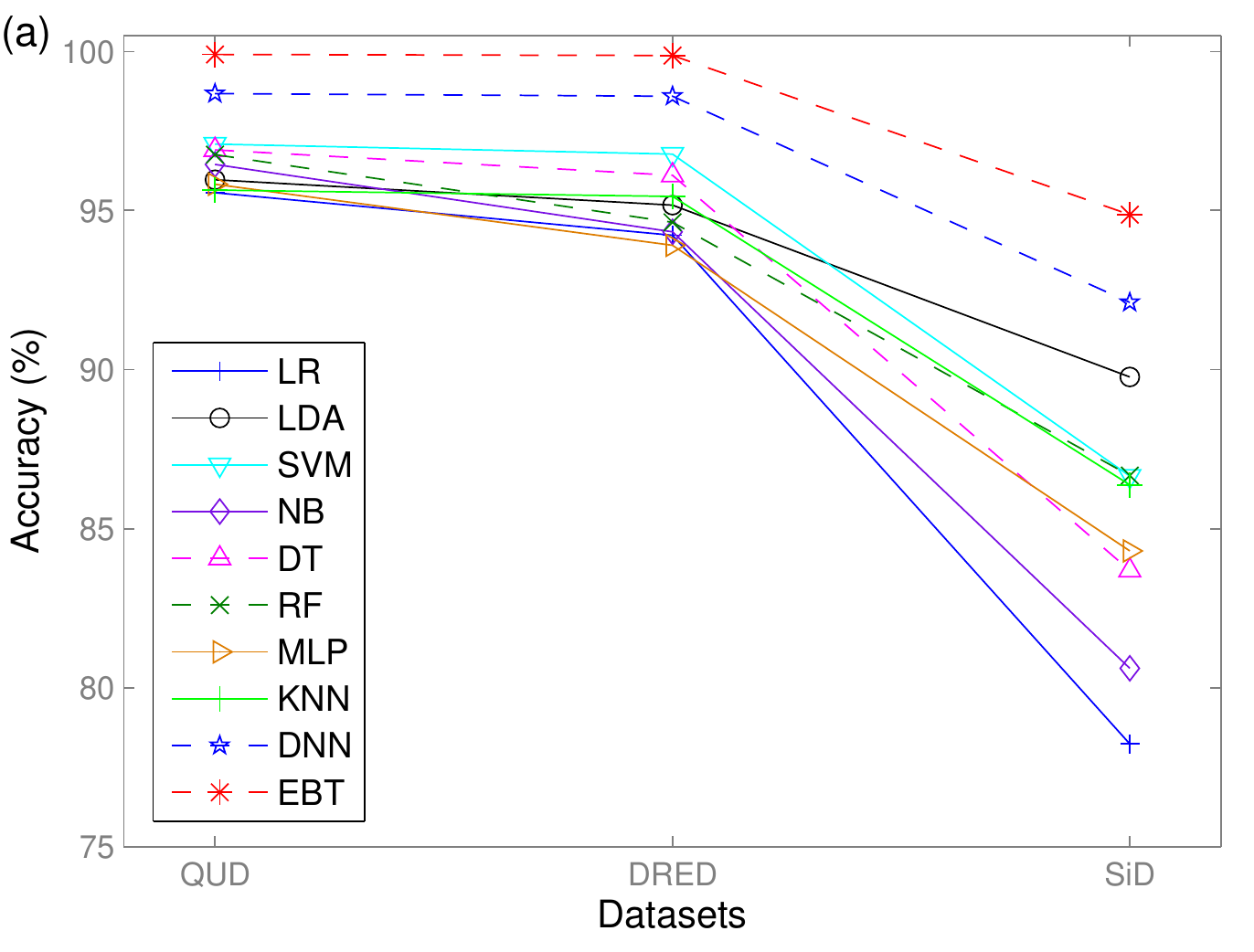}
\includegraphics[width=0.45\textwidth]{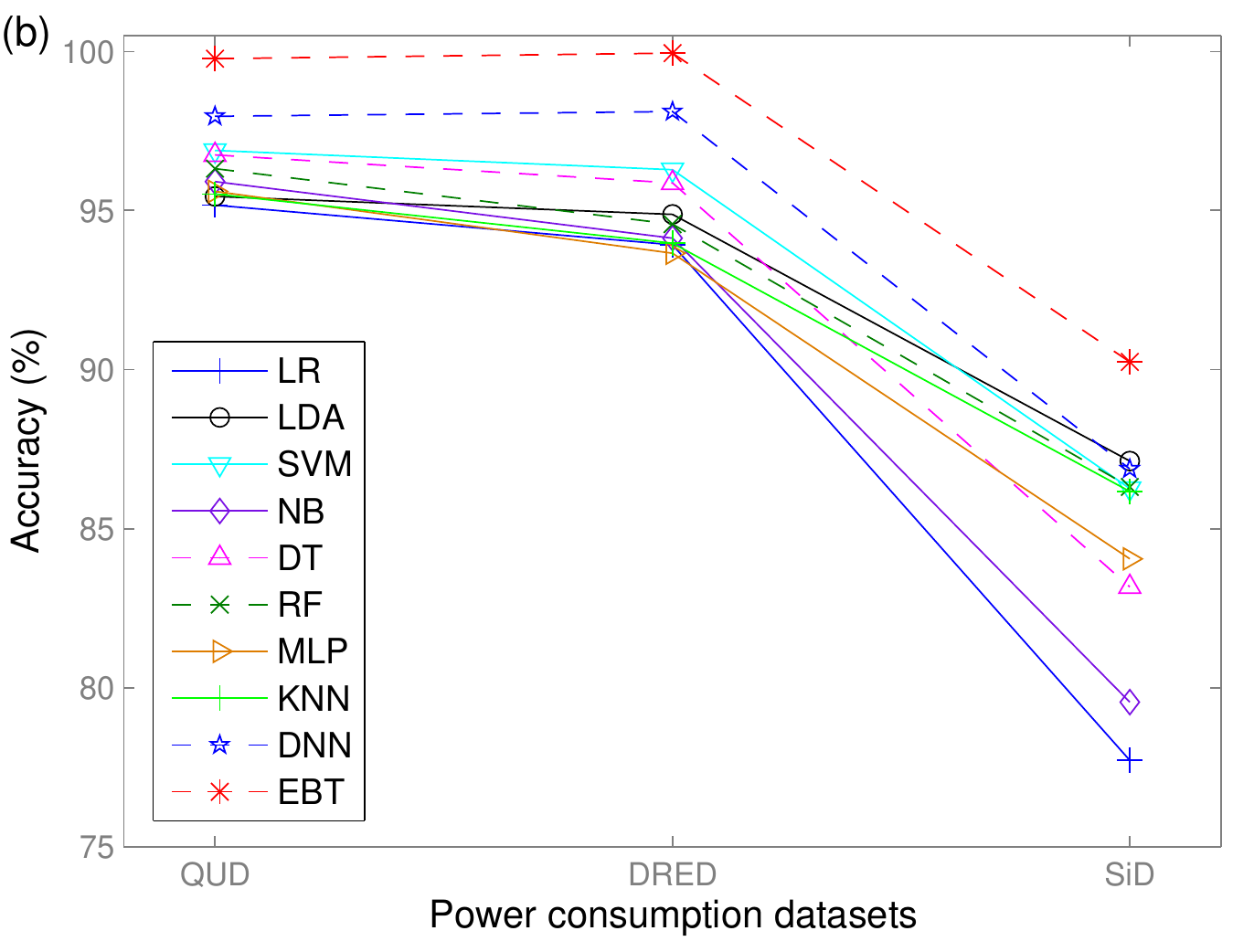}
\end{center}
\caption{Anomaly detection performance (a) accuracy, and (b) F1 score achieved under QUD, DRED, and SiD datasets.}
\label{ACC-F1}
\end{figure}

Fig. \ref{ACC-F1} illustrates the obtained results with reference to (a) accuracy, and (b) F1 score. \textcolor{black}{Indeed, using only the accuracy referring to the percentage of the correctly identified anomalies is not considered robust when dealing with imbalanced datasets (there is an unequal distribution of patterns over the classes), which is the case of the micro-moment data collected in this work. Therefore, the F1 score, which is more convenient to this scenario, has been employed as well for guaranteeing a fair performance assessment}. It has been clearly seen that the proposed EBT classifier outperforms the other classifiers under the three datasets. Accordingly, using EBT, more than 1.2\%, 1.3\%, and 2.7\% accuracy improvement have been obtained compared to DNN (which is ranked in the second position) under QUD, DRED, and SiD, respectively. On the other hand, more than 1.8\%, 1.85\%, and 3.3\% F1 score improvements have been achieved compared to DNN under QUD, DRED, and SiD, respectively.

\subsection{RS scenarios}
\begin{figure}[!t]
\centering
\begin{subfigure}{.15\textwidth}
  \centering
  \includegraphics[width=0.85\linewidth]{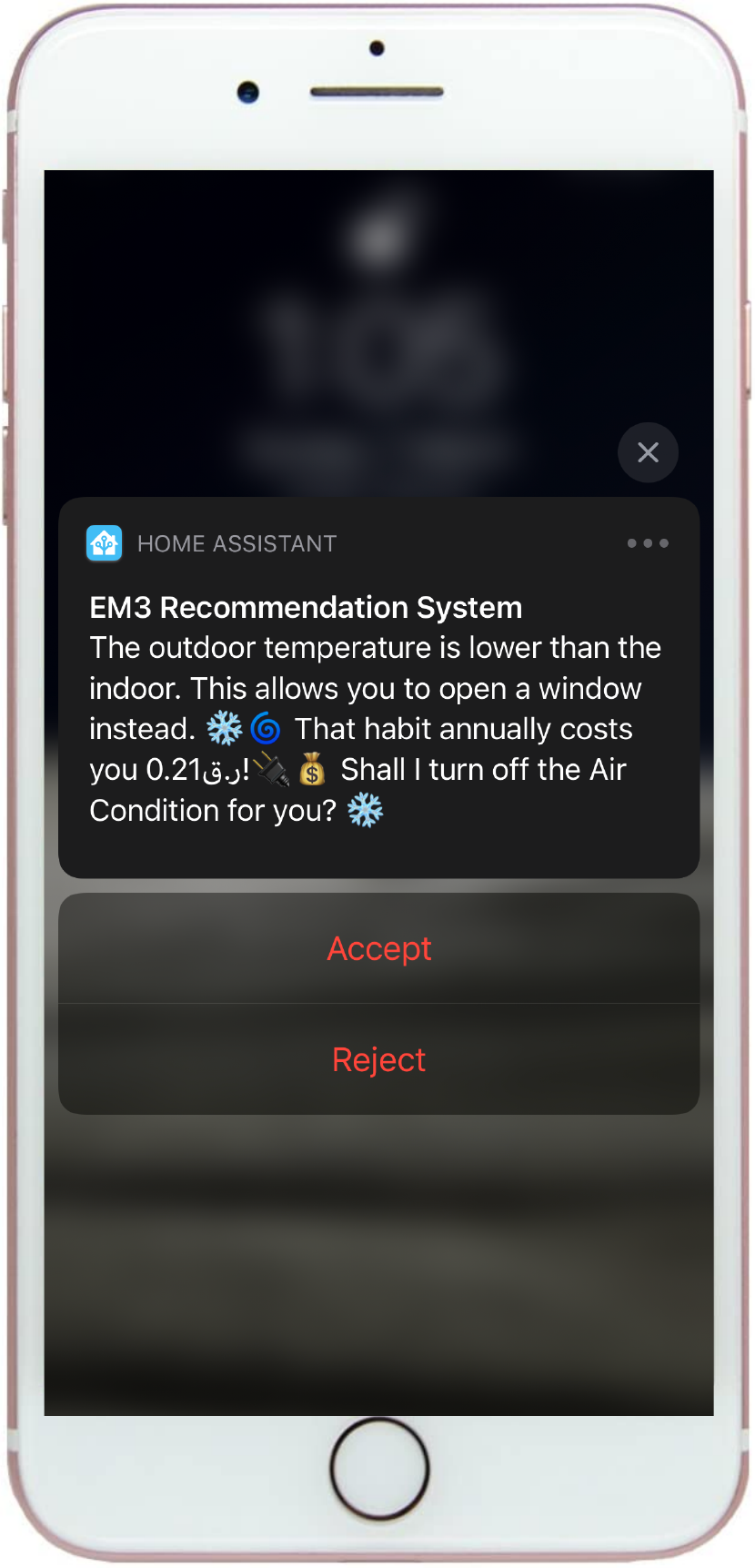}  
  \caption{}
  \label{fig:sub-first-notification}
\end{subfigure}
\begin{subfigure}{.15\textwidth}
  \centering
  \includegraphics[width=0.85\linewidth]{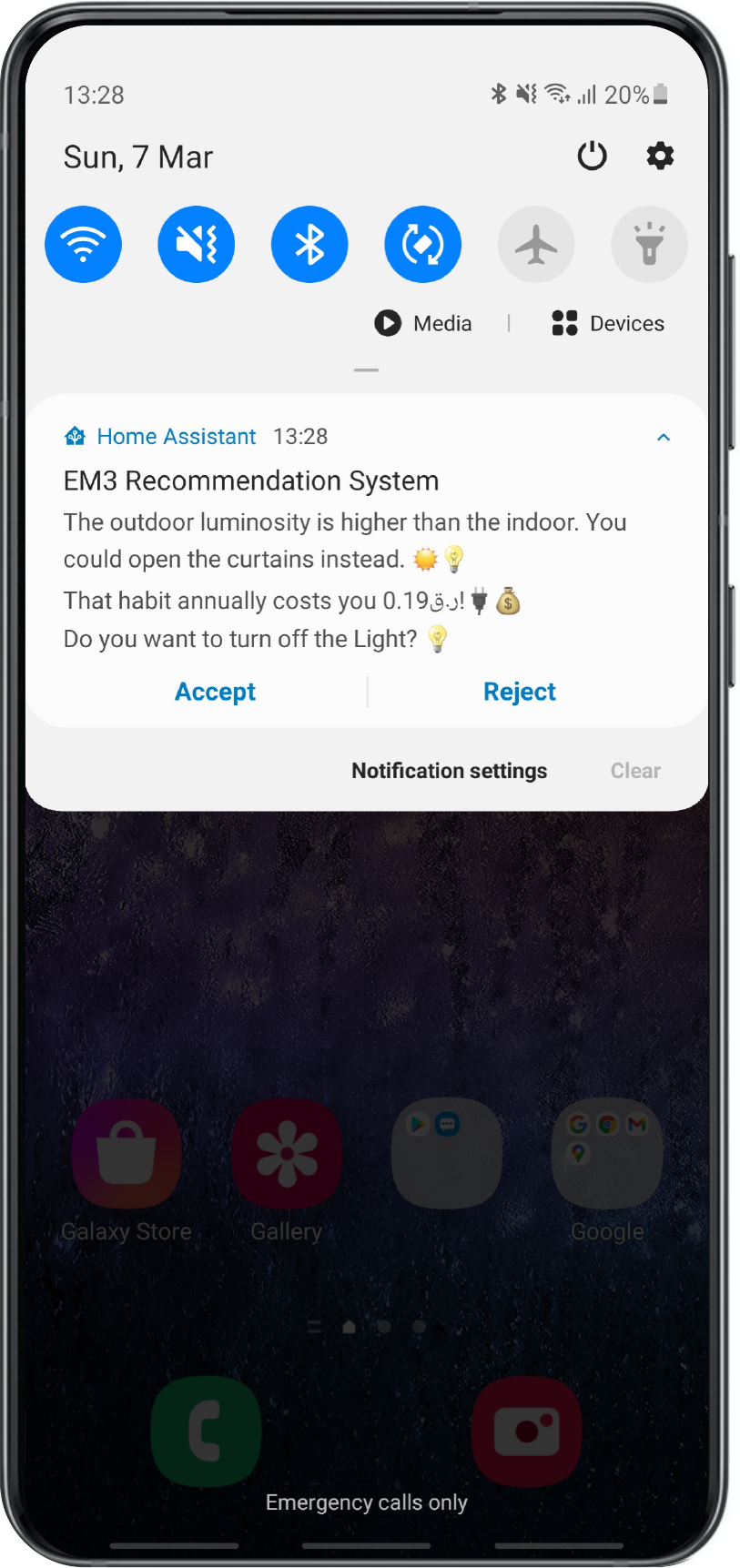}  
  \caption{}
  \label{fig:sub-second-notification}
\end{subfigure}
\begin{subfigure}{.15\textwidth}
  \centering
  \includegraphics[width=0.85\linewidth]{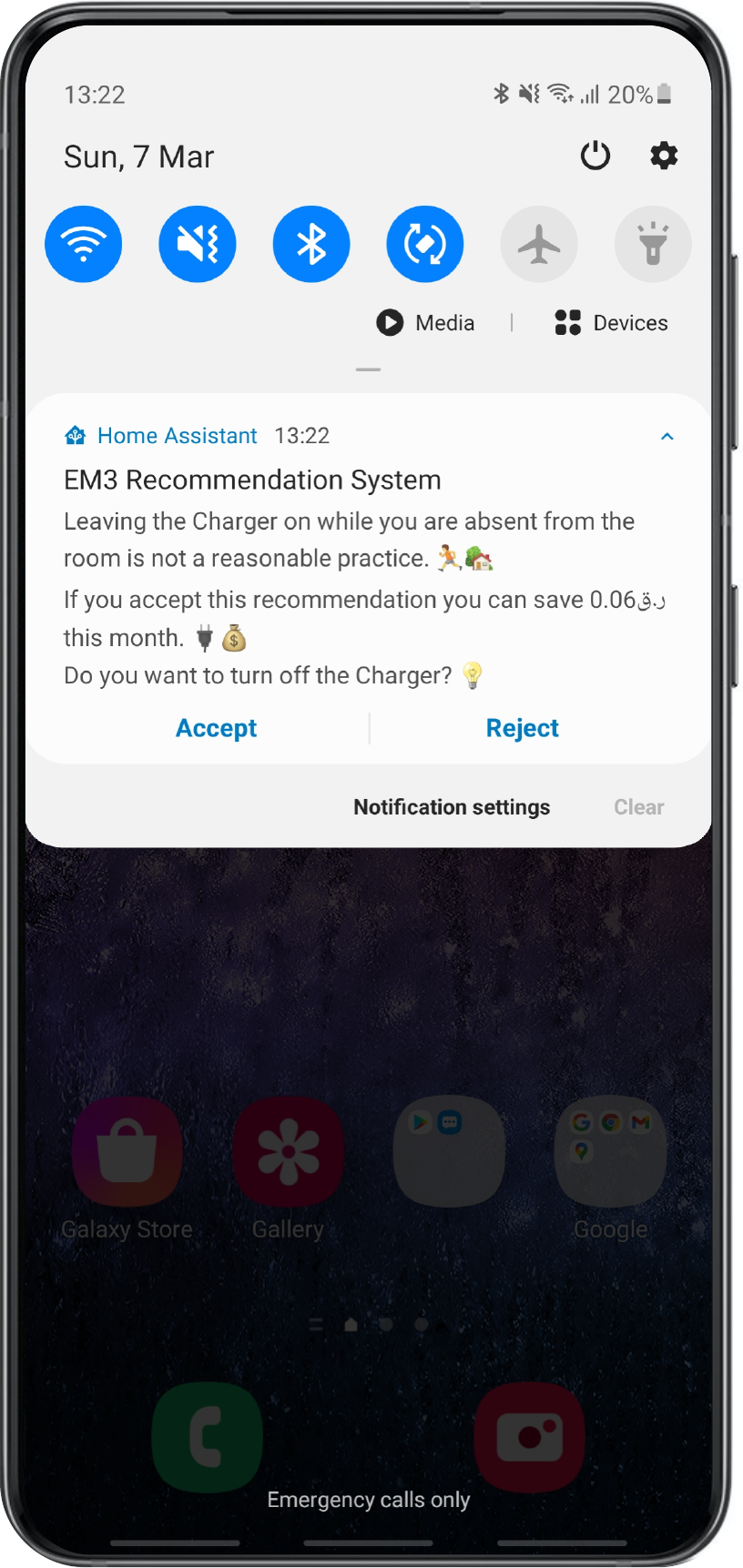}  
  \caption{}
  \label{fig:sub-third-notification}
\end{subfigure}

\caption{Pop-up recommendations presented to the user, when (a) an AC is excessively used, delivered to iOS device, (b) a light lamp is unnecessarily ON, and (c) an appliance is operating when the user is away, delivered to Android device.}
\label{em3-notification}
\end{figure}

\textcolor{black}{A real-world experiment was done by three research team users, where the system was installed in an academic building, as a proof-of-concept. However, since this framework includes many contributions, we could not elaborate on all of them in this study. Put simply, the experiment was done merely to ensure the overall flow of the framework and evaluate the performance of some subsystems.}
Therefore, the following scenarios were used to evaluate the influence of the recommendations in a single user's energy footprint: i) excessive Air Conditioner (AC) usage if the temperature outside is lower than inside and if the user is out of the room, ii) excessive light usage, if there is a sufficient amount of natural light and if the user is out of the room, and iii) excessive appliance usages when the user is out of the room. The advice is offered to the end-user with a justification for the trigger, an incentive to conserve energy and the choice to accept, reject, or ignore it. Moreover, the system is operational on either an Android or iOS supported mobile device. The three scenarios are shown in Fig. \ref{em3-notification} on an Android/iOS supported device, with the first scenario duplicated on an iOS device to illustrate system compatibility. The recommendations delivered to the user are conveyed in an explainable and persuadable manner, as seen in the preceding figure, to increase the user's willingness to change. \textcolor{black}{Typically, Fig. \ref{em3-notification} (a) depicts the notification sent to the user to indicate that the outdoor temperature is lower than the indoor and ask him/her to open the window instead of using the AC. In Fig.  \ref{em3-notification} (b), the user has been notified that the outdoor luminosity is higher than the indoor, and has been asked to open the curtains instead of using the light. In Fig.  \ref{em3-notification} (c), the user has been informed that the charger is on while he/she is absent, and has been invited to turn it off. Moreover, for each provided recommendation, an estimation of the actual saving cost (in Qatari Riyal) is included for pursuing the user to accept it. }  

depicted in Figure 5. Recommendations are composed of two different sections, which serve the explainable and persuasive recommendations scenario, respectively. The messages contain visual cues that facilitate users to understand the reason of each recommendation.

\subsection{Comparison with prior works}
Although a great deal of attention is devoted to developing RSs, less attention has been put to design reproduction toolkits platforms, hence, the reproducibility of existing RSs is still an issue and experimental comparison is often difficult even impossible. To summarize, the main issues that impede a fair comparison between state-of-the-art RSs are (i) absence of specific datasets to test the RSs since most of the frameworks use their own data, (ii) data preprocessing that is used to correct uncompleted data and/or resample recorded data, which is usually made with reference to particular needs of the users, kinds of devices observed during the evaluation and monitoring period, and (iii) diversity of the evaluation metrics used to assess the performance of RSs, which makes the comparison an arduous task.
Using what we have at hand, the proposed RS has been compared with various state-of-the-art RSs. Accordingly, Table \ref{comp-tab} outlines a comparison in terms of the use of explanations, cloud computing, edge computing, real-time or near real-time implementation, and privacy preservation. In addition to using explanations to increase the users' trust and the recommendation acceptance rate, this framework is the first energy RS that has been run on edge devices. Therefore, it presents better privacy preservation potential than those implemented on cloud platforms since running a RS on edge devices helps overcome the privacy problems corresponding to the transmission and storage of sensitive data on the cloud. \textcolor{black}{
Moreover, the edge can deal with the data in an intelligent way by implementing the machine learning algorithms used for analyzing energy data and detecting anomalous energy consumption on an edge server. Put simply, the intelligence and computing tasks have been brought to the edge device, which helps in reducing the latency and avoiding the bandwidth constraints and Internet dependency, by contrast to using cloud.}


\begin{table}[!t]
\centering
\caption{Comparison of the proposed edge-based RS with existing RSs with reference to different characteristics. }
\label{comp-tab}

\begin{tabular}{lllll}
\hline
{\small Work} & {\small Explanations} & {\small Computing} & {\small %
Real-time} & {\small Privacy } \\ 
&  & {\small platform} & {\small near real-time} & {\small preservation } \\ 
\hline
{\small \cite{starke2015saving}} & $\times $ & {\small N/A} & $\times $ & $%
\times $ \\ 
{\small \cite{starke2020little}} & ${\small \checkmark }$ & {\small cloud} & 
$\times $ & $\times $ \\ 
{\small \cite{sardianos2020rehab}} & $\times $ & {\small cloud} & $\times $
& $\times $ \\ 
{\small \cite{wei2020deep}} & $\times $ & {\small cloud} & $\times $ & $%
\times $ \\ 
{\small \cite{sardianos2021emergence}} & ${\small \checkmark }$ & {\small %
cloud} & ${\small \checkmark }$ & $\times $ \\ 
{\small \cite{zhang2019bayesian} \ } & $\times $ & {\small cloud} & ${\small %
\checkmark }$ & $\times $ \\ 
{\small Our} & ${\small \checkmark }$ & {\small edge} & ${\small \checkmark }
$ & ${\small \checkmark }$ \\ \hline
\end{tabular}

\end{table}

\section{Conclusion} \label{sec6}
In this paper, we present an implementation of a real-case scenario of our energy management system within Qatar University campus. Current findings suggest that by exploiting the advantages of an explainable recommender system, an open-source platform like Home-Assistant with the installation of sensors, smart plugs, and a consumption monitoring system, we are able to promote the users' transition to new consumption patterns. It will be part of our future work to make more efforts to improve the framework, including evaluating the system performing with multiple users by expanding the number of sensors and smart plugs employed. Additionally, we propose adding more data visualization options as a persuasive method and extending optimization methods to enhance the overall energy-saving performance. Moving on, the use of different edge architectures and computing modes to improve the performance of edge computing will be investigated as well, such as binary offloading, model partitioning, and distributed computation. 

\textcolor{black}{
Finally, it is worth noting that further investigations are required by extending the number of participants to come up with a detailed evaluation of the whole system, which will be part of our future work. Accordingly, a mini-pilot survey will be conducted to identify the energy-saving characteristics of participants and provide foresight to predict how consumers can save energy on a daily basis. For the mini-pilot, a survey will be given in the beginning and another at the end to observe any change in answers after being involved in such an energy-saving study.}

\section*{Acknowledgements}
This paper was made possible by National Priorities Research Program (NPRP) grant No. 10-0130-170288 from the Qatar National Research Fund (a member of Qatar Foundation). The statements made herein are solely the responsibility of the authors. 



\end{document}